\documentclass[12pt]{article}

\usepackage[dvipdfmx]{graphicx}
\usepackage{amsfonts}
\usepackage[mathscr]{eucal}
\usepackage{ascmac}
\usepackage{dcolumn}
\usepackage{bm}
\usepackage[colorlinks=true,linkcolor=blue,citecolor=blue]{hyperref}
\usepackage{hyperref}
\usepackage{color}

\begin{document}


\newcommand{\gtrsim}{ \mathop{}_{\textstyle \sim}^{\textstyle >} }
\newcommand{\lesssim}{ \mathop{}_{\textstyle \sim}^{\textstyle <} }
\newcommand{\vev}[1]{ \left\langle {#1} \right\rangle }
\newcommand{\bra}[1]{ \langle {#1} | }
\newcommand{\ket}[1]{ | {#1} \rangle }
\newcommand{\EV}{ \ {\rm eV} }
\newcommand{\KEV}{ \ {\rm keV} }
\newcommand{\MEV}{\  {\rm MeV} }
\newcommand{\GEV}{\  {\rm GeV} }
\newcommand{\TEV}{\  {\rm TeV} }
\newcommand{\1}{\mbox{1}\hspace{-0.25em}\mbox{l}}
\newcommand{\Red}[1]{{\color{red} {#1}}}

\newcommand{\lmk}{\left(}  
\newcommand{\rmk}{\right)}
\newcommand{\lkk}{\left[}  
\newcommand{\rkk}{\right]}
\newcommand{\lhk}{\left \{ }  
\newcommand{\rhk}{\right \} }
\newcommand{\del}{\partial}  
\newcommand{\la}{\left\langle} 
\newcommand{\ra}{\right\rangle}
\newcommand{\half}{\frac{1}{2}}

\newcommand{\bea}{\begin{array}}
\newcommand{\eea}{\end{array}}
\newcommand{\beq}{\begin{eqnarray}}
\newcommand{\eeq}{\end{eqnarray}}

\newcommand{\dd}{\mathrm{d}}
\newcommand{\Mpl}{M_{\rm Pl}}
\newcommand{\mg}{m_{3/2}}
\newcommand{\abs}[1]{\left\vert {#1} \right\vert}
\newcommand{\mphi}{m_{\phi}}
\newcommand{\Hz}{\ {\rm Hz}}
\newcommand{\for}{\quad \text{for }}
\newcommand{\Min}{\text{Min}}
\newcommand{\Max}{\text{Max}}
\newcommand{\Kahler}{K\"{a}hler }
\newcommand{\cphi}{\varphi}
\newcommand{\Tr}{\text{Tr}}
\newcommand{\diag}{\text{diag}}

\newcommand{\SUf}{SU(3)_{\rm f}}
\newcommand{\Upq}{U(1)_{\rm PQ}}
\newcommand{\Zpq}{Z^{\rm PQ}_3}
\newcommand{\Cpq}{C_{\rm PQ}}
\newcommand{\ubar}{u^c}
\newcommand{\dbar}{d^c}
\newcommand{\ebar}{e^c}
\newcommand{\nubar}{\nu^c}
\newcommand{\Ndw}{N_{\rm DW}}
\newcommand{\Fpq}{F_{\rm PQ}}
\newcommand{\fpq}{v_{\rm PQ}}
\newcommand{\Br}{{\rm Br}}
\newcommand{\Lag}{\mathcal{L}}
\newcommand{\Lqcd}{\Lambda_{\rm QCD}}


\begin{titlepage}

\baselineskip 8mm

\begin{flushright}
IPMU 15-0049
\end{flushright}

\begin{center}

\vskip 1.2cm

{\Large\bf
Observable~dark~radiation from~cosmologically~safe~QCD~axion
}

\vskip 1.8cm

{\large Masahiro Kawasaki$^{a,b}$, 
Masaki Yamada$^{a,b}$, 
and 
Tsutomu T. Yanagida$^{b}$
}

\vskip 0.4cm

{\it$^a$Institute for Cosmic Ray Research, The University of Tokyo,
Kashiwa, Chiba 277-8582, Japan}\\
{\it$^b$Kavli IPMU (WPI), TODIAS, The University of Tokyo, 
Kashiwa, Chiba 277-8583, Japan}

\date{\today}
\vspace{2cm}

\begin{abstract}  
We propose a QCD axion model that avoids the cosmological domain wall problem, introducing a global $SU(3)_{\rm f}$ family symmetry to which we embed the unwanted PQ discrete symmetry. The spontaneous breaking of $SU(3)_{\rm f}$ and PQ symmetry predicts eight NG bosons as well as axion, all of which contribute to dark radiation in the Universe. The derivation from the standard model prediction of dark radiation can be observed by future observations of CMB fluctuations. Our model also predicts a sizable exotic kaon decay rate, which is marginally consistent with the present collider data and would be tested by future collider experiments. 
\end{abstract}


\end{center}
\end{titlepage}

\baselineskip 6mm


\section{Introduction
\label{sec:introduction}}

After the discovery of the $126 \GEV$ Higgs boson at the LHC~\cite{Aad:2012tfa, Chatrchyan:2012ufa}, 
we obtain the complete parameter set of the Standard Model (SM) of particle physics 
except for the remaining parameter, the strong CP phase~\cite{Callan:1976je, Jackiw:1976pf}. 
We only know the upper bound on the strong CP phase, $\abs{\theta_{\rm QCD}} \lesssim 10^{-(10-11)}$, 
which comes from the experimental upper bound on the neutron electric dipole moment~\cite{Baker:2006ts}. 
This is unnaturally small, 
and thus we face a severe fine tuning problem called the strong CP problem. 
In addition to that, the observations of neutrino oscillations require new physics beyond the SM, 
and the origins of baryon asymmetry and dark matter (DM) 
are also long-standing challenges in cosmology and particle physics. 
One of the simplest solutions to those problems is a QCD axion model~\cite{Peccei:1977hh, Peccei:1977ur, Kim:1979if, Dine:1981rt} 
with right-handed neutrinos~\cite{seesaw, Fukugita:1986hr}. 
The strong CP phase is cancelled by the vacuum expectation value (VEV) of axion~\cite{Weinberg:1977ma}, 
which is the pseudo-NG boson 
associated with the spontaneous symmetry breaking (SSB) of Peccei-Quinn (PQ) symmetry. 
The axion can be generated as a coherent oscillation in the early Universe 
and is a good candidate of cold DM~\cite{Preskill:1982cy, Abbott:1982af, Dine:1982ah}. 
Introducing heavy right-handed neutrinos, 
we can explain small neutrino masses via the seesaw mechanism. 
The out of equilibrium decay of the right-handed neutrinos can generate lepton asymmetry, 
which is then converted to baryon asymmetry via the sphaleron effect~\cite{Kuzmin:1985mm}.

The above simple scenario may confront the problem of 
sizable isocurvature fluctuations~\cite{Axenides:1983hj, Seckel:1985tj, Turner:1990uz} 
or the cosmological domain wall problem~\cite{Zeldovich:1974uw, Sikivie:1982qv}. 
If the PQ symmetry is spontaneously broken before inflation, 
it predicts a sizable isocurvature density perturbations due to quantum fluctuations of the axion field during inflation. 
The resulting amount of isocurvature perturbations is inconsistent with the observation of CMB fluctuations 
unless the energy scale of inflation is relatively small to suppress the amplitude of axion fluctuations. 
On the other hand, 
it is possible that the PQ symmetry is broken after inflation. 
In this case, 
the SSB of PQ symmetry results in formation of cosmic strings, 
and then at the QCD phase transition 
domain walls whose boundaries are cosmic strings are formed~\cite{Vilenkin:1982ks}.%
\footnote{%
Even if the PQ symmetry is broken during inflation, 
the PQ symmetry may be restored due to the thermal effects 
after the reheating for sufficiently high reheating temperature 
suggested from the successful leptogenesis~\cite{Buchmuller:2005eh}. 
}
The domain walls are stable and soon dominate the Universe, 
so that they spoil the success of the standard cosmological scenario. 
Although 
this is not the case if the domain wall number is equal to one, 
it restricts axion models. 
In the DFSZ axion model~\cite{Dine:1981rt}, in particular, 
the domain wall number is equal to 6, so that it confronts the domain wall problem. 

In Ref.~\cite{Lazarides:1982tw}, Lazarides and Shafi have proposed a novel mechanism 
that solves the domain wall problem. 
They introduced an additional continuous symmetry to which the unwanted discrete symmetry is embedded 
as a center of the continuous group. 
Since the degenerate vacua is connected with each other through the continuous symmetry, 
there is no domain wall problem in this case.

In this paper, we consider a variant of the DFSZ axion model 
and introduce a global $\SUf$ flavour symmetry for three families of quarks and leptons 
to realize the Lazarides-Shafi mechanism. 
The unwanted discrete PQ symmetry is embedded to the $\SUf$ symmetry 
so that the vacua are continuously connected to avoid the formation of stable domain walls. 
Both $\SUf$ and PQ symmetry are spontaneously broken at the same time, 
so that the low energy effective theory contains eight NG bosons, called familons~\cite{Wilczek:1982rv, 
Reiss:1982sq, Gelmini:1982zz, Kim:1986ax}, as well as the axion. 
We find that they are thermalized and then decoupled from the thermal plasma after the SSB, 
and give a sizable contribution to dark radiation 
in the subsequent cosmological history~\cite{Nakayama:2010vs, Weinberg:2013kea}. 
We find that the resulting amount of the dark radiation is consistent with the present 
observed value of effective neutrino number, 
and the discrepancy from the standard model prediction is measurable by future observations of CMB fluctuations. 
In addition, the SSB of the $\SUf$ symmetry predicts collider signatures in terms of flavour changing processes. 
In particular, the exotic kaon decay rate is marginally consistent with the present constraint 
and would be measured by future collider experiments.

In the next section, 
we introduce a variant of the DFSZ axion model with a global $\SUf$ flavour symmetry 
for three quarks and leptons, and briefly explain the Lazarides-Shafi mechanism. 
In Sec.~\ref{sec:predictions}, we consider constraints and predictions of our model: 
cold axion DM, dark radiation, exotic kaon decay, and baryon asymmetry. 
Section~\ref{sec:conclusion} is devoted to the conclusion. 

\section{Model
\label{sec:model}}

We consider a variant of the DFSZ model with global $\SUf \times \Upq$ symmetry. 
The charge assignment is shown in Table~\ref{table1}, 
where $\chi$ is the field responsible for the SSB of $\SUf \times \Upq$. 
The right-handed neutrinos $\nubar$ are introduced to realize the seesaw mechanism~\cite{seesaw} 
and thermal leptogenesis~\cite{Fukugita:1986hr}, 
which we discuss in Sec.~\ref{subsec:leptogenesis}. 
Note that the $\SUf$ symmetry has anomaly, so that it has to be a global symmetry. 
The $\SUf \times \Upq$ symmetry is spontaneously broken completely by the VEV of $\chi$, 
which we denote as 
\beq
 \la \chi \ra^{ij} = \frac{1}{\sqrt{2}} {\rm diag} \lmk v_1, v_2, v_3 \rmk. 
\eeq
The SSB predicts eight NG bosons 
called familons~\cite{Wilczek:1982rv, Reiss:1982sq, Gelmini:1982zz, Kim:1986ax} 
as well as pseudo-NG boson called axion~\cite{Weinberg:1977ma} 
in the low energy effective theory.

\begin{table}\begin{center}
\begin{tabular}{|c|p{1.0cm}|p{1.0cm}|p{1.0cm}|p{1.0cm}|p{1.0cm}|p{1.0cm}|p{1.0cm}|p{1.0cm}|p{1.0cm}|p{1.0cm}|}
  \hline
    & \hfil $Q$ \hfil & \hfil $\ubar$ \hfil & \hfil $\dbar$ \hfil & \hfil $L$ \hfil & \hfil $\ebar$ \hfil & \hfil $\nubar$  \hfil & \hfil $H_u$ \hfil & \hfil $H_d$\hfil & \hfil $\chi$\hfil \\
  \hline
  \hfil $\SUf$ \hfil & \hfil {\bf 3} \hfil & \hfil {\bf 3} \hfil & \hfil {\bf 3} \hfil & \hfil {\bf 3} \hfil & \hfil {\bf 3} \hfil & \hfil {\bf 3} \hfil & \hfil ${\bf 6}^*$ \hfil & \hfil ${\bf 6}^*$ \hfil & \hfil ${\bf 6}^*$ \hfil  \\
  \hline
  \hfil $U(1)_{\rm PQ}$ \hfil & \hfil $1/4$ \hfil & \hfil $1/4$ \hfil & \hfil $1/4$ \hfil & \hfil $1/4$ \hfil & \hfil $1/4$ \hfil & \hfil $1/4$ \hfil & \hfil $-1/2$ \hfil & \hfil $-1/2$ \hfil & \hfil $1$ \hfil \\
\hline
\end{tabular}\end{center}
\caption{Charge assignment for matter fields.
\label{table1}}
\end{table}

The axion couples with the gluon field through the anomalous interaction: 
\beq
 \frac{g_s^2}{32 \pi^2} \frac{a}{F_a} G^{a \mu \nu} \tilde{G}_{\mu \nu}^a, 
\eeq
where $g_s$ is the QCD gauge coupling constant, 
$G^{a \mu \nu}$ is the gluon field strength, 
and $\tilde{G}_{\mu \nu}^a$ is its dual. 
The axion decay constant $F_a$ is given as 
\beq
 \frac{1}{F_a^2} \simeq \lkk \frac{1}{v_1^2} + \frac{1}{v_2^2} + \frac{1}{v_3^2} \rkk. 
 \label{F_a}
\eeq

In this paper, we consider the case that 
the $\SUf \times \Upq$ symmetry is spontaneously broken after inflation. 
Since the first homotopy group is given by $\pi_1 \lkk \SUf \times \Upq \rkk = Z$, 
cosmic strings form at the time of the SSB of the $\SUf \times \Upq$ symmetry~\cite{Vilenkin:1982ks}. 
Then, after the QCD phase transition, 
the axion field effectively acquires a periodic potential through the QCD instanton effect. 
The non-perturbative QCD effects associated with instantons 
break $\Upq$ to $Z_3^{\rm PQ}$ in our model~\cite{Sikivie:1982qv, 'tHooft:1976up, 'tHooft:1976fv}, 
so that domain walls may appear as bounded by the cosmic strings at the QCD phase transition.

We define the domain wall number $\Ndw$ as the number of disconnected vacua. 
Each cosmic string becomes the boundary of $\Ndw$ domain walls after the QCD phase transition. 
If the $\SUf$ symmetry was absent, 
it would be given by $\Ndw = 3$ because the non-perturbative QCD effects break $\Upq$ to $Z_{3}^{\rm PQ}$. 
A domain wall system with $\Ndw \ge 2$ is stable and thus eventually dominate the Universe, 
in which case the subsequent cosmological history 
is highly inconsistent with the observations~\cite{Zeldovich:1974uw, Sikivie:1982qv}. 
However, 
we introduce the $\SUf$ symmetry to embed the $Z_3^{\rm PQ}$ symmetry into the continuous symmetry, 
so that the degenerate vacua are continuously connected with each other.%
\footnote{%
In Ref.~\cite{Barr:1982bb} they consider a GUT theory with $SO(10) \times \SUf \times U(1)_{\rm PQ}$ to realize the Lazarides-Shafi mechanism. 
They use the center of $SO(10) \times \SUf$ to embed the $Z_{6}^{\rm PQ}$ discrete symmetry into the continuous group. 
In this model, however, 
the PQ symmetry breaking as well as the $\SUf$ symmetry breaking occur at the GUT scale. 
}
This implies that 
there is only one minimum along the axion field, 
so that each cosmic string is attached by only one domain wall ($N_{\rm DW} = 1$) 
after the axion acquires the effective mass at the QCD phase transition. 
In this case, the domain wall is short lived due to their tension, 
and the subsequent cosmological scenario can be consistent with the standard cosmology~\cite{Vilenkin:1982ks, Lazarides:1982tw}.

Here, let us check that the $Z_3^{\rm PQ}$ symmetry is actually embedded to the $\SUf$ symmetry. 
The $\Upq$ symmetry commutes with $\SUf$, so that its discrete subgroup $Z_3^{\rm PQ}$ also commutes with it. 
Therefore, what we should check is that the $Z_3^{\rm PQ}$ group is embedded to the center of $\SUf$ 
because the center of a group commutes with all the group elements by definition. 
The center of $\SUf$ group is $Z_3$ and its effect on matter fields can be written as 
\beq
 Q &\to& e^{2 \pi i (n/3)} Q \\
 H_u &\to& e^{2 \pi i(-2n/3)} H_u \\
 \chi &\to& e^{i 2 \pi (-2n/3)} \chi = e^{i 2 \pi (4n/3)} \chi, 
\eeq
where $n$ ($=1,2,3$) is an integer. 
The other matter fields are also transformed in a similar way. 
Here, you can see that 
the center of $\SUf$ transforms the matter fields 
in the same way as $\Upq$ with an order parameter of $\theta = 2\pi (n/3)$, 
which is indeed the $Z_3^{\rm PQ}$ discrete subgroup of $\Upq$ transformation. 
This means that the $Z_3^{\rm PQ}$ group is identified with the center of $\SUf$, 
and thus it is embedded to the continuous $\SUf$ group. 
This implies that the degenerate $N_f$ vacua are continuously connected with each other by the $\SUf$ transformation.

Here, we comment on the general case in which the number of family is $N_f$. 
Let us consider a variant of the DFSZ axion model 
where the interaction term between the Higgs fields and SSB breaking field $\chi$ 
is given by $H_u H_d \chi^n$ with a certain integer $n$. 
In this case, the non-perturbative QCD effects associated with instantons break $\Upq$ to $Z_{n N_f}^{\rm PQ}$. 
Here, 
we need to embed $Z_{n N_f}^{\rm PQ}$ symmetry to the $SU(N_f)$ symmetry 
to solve the domain wall problem by the Lazarides-Shafi mechanism. 
However, 
the center of $SU(N_f)$ is $Z_{N_f}$, 
so that $n$ has to be equal to unity. 
Then, we should check wether 
the interaction term $H_u H_d \chi$ is consistent with the $SU(N_f)$ symmetry. 
In the case of $N_f = 3$ it is actually consistent, 
while in the other cases it may not be. 
This implies that 
only the case with $N_f = 3$ can realize the Lazarides-Shafi mechanism to solve the domain wall problem. 
Although the case with $N_f = 1$ is also no domain wall problem, 
it is disfavoured in light of leptogenesis since there is no CP-violating phase. 
Therefore, our scenario may explain that 
the number of family is equal to $3$.

\section{Constraints and predictions 
\label{sec:predictions}}

In this section, we consider constraints and predictions in our model. 
Cold axions are generated around the time of the QCD phase transition from three mechanisms: 
misalignment mechanism~\cite{Preskill:1982cy, Abbott:1982af, Dine:1982ah}, 
emission from cosmic strings~\cite{Davis:1986xc}, and decay of domain walls~\cite{Lyth:1991bb}. 
The resulting axion abundance can be consistent with the observed DM abundance as shown in the next subsection. 
In Sec.~\ref{subsec:dark radiation}, 
we estimate the axion and familon decoupling temperature 
and show that they contribute to dark radiation in the Universe~\cite{Nakayama:2010vs, Weinberg:2013kea}. 
The resulting amount of dark radiation can be measured by future CMB observations. 
Then, 
in Sec.~\ref{subsec:collider signatures}, 
we calculate an exotic kaon decay rate that is severely constrained by collider experiments. 
The result is marginally consistent with the present constraint 
and would be measured by future collider experiments. 
Finally, we check that the seesaw mechanism and leptogenesis can be naturally realized in our model.

\subsection{Axion dark matter
\label{subsec:DM}}

In the case that the $\Upq$ symmetry is spontaneously broken after inflation as we considered in this paper, 
cold axions are generated from three mechanisms: 
the misalignment mechanism~\cite{Preskill:1982cy, Abbott:1982af, Dine:1982ah}, 
emission from cosmic strings~\cite{Davis:1986xc}, and decay of domain walls~\cite{Lyth:1991bb}.%
\footnote{%
A typical momentum of axions generated from topological defects is of the order of the Hubble parameter, 
so that these axions soon become nonrelativistic due to the redshift effect and can be cold DM~\cite{Kawasaki:2014sqa}. 
}
The resulting amounts of axions from these mechanisms are the same order with each other. 
The detailed calculations are performed in Ref.~\cite{Kawasaki:2014sqa}, 
though we need some miner corrections to apply the results to our model.

First, let us explain the axion production from the misalignment mechanism~\cite{Preskill:1982cy, Abbott:1982af, Dine:1982ah}. 
At a temperature above the QCD phase transition, 
the axion field stays at phases that are randomly distributed over the horizon scale. 
As the temperature decreases, it acquires the effective mass through the nonperturbative effect~\cite{'tHooft:1976up, 
'tHooft:1976fv}. 
A study based on the interacting instanton liquid model 
implies that the axion mass in a finite temperature can be fitted by the following power law formula~\cite{Wantz:2009it}: 
\beq
 m_a (T)^2 = c_T \frac{\Lambda_{\rm QCD}^4}{F_a^2} \lmk \frac{T}{\Lqcd} \rmk^{-\alpha}, 
\eeq
where $c_T = 1.68 \times 10^{-7}$, $\alpha=6.68$, and $\Lqcd = 400 \MEV$. 
This power law formula should be truncated 
when the axion mass reaches the zero temperature value of 
\beq
 m_a (0) \simeq \frac{m_u m_d}{\lmk m_u + m_d \rmk^2} \frac{m_\pi f_\pi}{F_a}. 
\eeq
When the Hubble parameter becomes to be comparable to the axion mass, 
the axion field starts to oscillate around the minimum of the potential. 
It occurs at the time around $t=t_{\rm osc}$ that is defined by $m_a(t_{\rm osc}) = H(t_{\rm osc})$. 
Since the axion field oscillation is adiabatic, its number density is conserved after the oscillation begins. 
Thus we obtain the axion abundance generated from the misalignment mechanism as 
\beq
 \Omega_a^{\rm mis} h^2 
 \simeq 
 (0.07 -0.09)
 \lmk \frac{F_a}{10^{11} \GEV} \rmk^{1.19} 
  \lmk \frac{\Lqcd}{400 \MEV} \rmk, 
\eeq
where we have substituted $\mathcal{O}(1)$ numerical parameters~\cite{Kawasaki:2014sqa} 
and $h$ is the Hubble parameter in units of $100 \ {\rm km/s/Mpc}$. 
Note that the result is independent of initial conditions 
because the initial phase of the axion field is randomly distributed 
and we should average it over the horizon scale to obtain the net axion abundance.

Next, we consider the axion emission from cosmic strings~\cite{Davis:1986xc}. 
Since we consider the case that the $\SUf \times \Upq$ symmetry is spontaneously broken after inflation, 
cosmic strings form at the time of SSB. 
The energy density per unit length of cosmic strings $\mu_{\rm string}$ is roughly given by 
\beq
 \mu_{\rm string} \simeq \pi F_a^2 \log \lmk \delta_s^{-1} t \rmk, 
\eeq
where $\delta_s$ is the core width of cosmic strings. 
Cosmic strings emit NG bosons (axions and familons) to follow the scaling dependence: 
\beq
 \rho_{\rm cs} \simeq \frac{\mu_{\rm string}}{t^2}, 
\eeq
where $\rho_{\rm cs}$ is the energy density of cosmic strings. 
Since cosmic strings lose their energy by emitting NG bosons, 
the energy density of NG bosons is roughly equal to that of cosmic strings. 
Here, the NG bosons consist of familons as well as axions, 
so that the energy density of axions have an additional $\mathcal{O}(1)$ uncertainty in our model. 
Integrating it from the time of PQ phase transition to $t_{\rm osc}$, 
we obtain the axion abundance from cosmic strings as 
\beq 
 \Omega_a^{\rm CS} h^2 = 
 (0.005 -0.2)
 \lmk \frac{F_a}{10^{11} \GEV} \rmk^{1.19} 
  \lmk \frac{\Lqcd}{400 \MEV} \rmk, 
\eeq
where we include uncertainties coming from numerical simulations. 
In addition to that, we conservatively include an uncertainty of $(1/9 - 1)$ coming from 
the efficiency of axion emission from cosmic strings, 
where the factor of $1/9$ corresponds to the case that 
cosmic strings emit all types of NG bosons equally.

Axions are also generated from the decay of domain walls~\cite{Lyth:1991bb}. 
Although the domain wall problem is solved by the Lazarides-Shafi mechanism in our model, 
short-lived domain walls with $\Ndw = 1$ form at the QCD phase transition ($t=t_{\rm osc}$). 
Since the domain walls are bounded by cosmic strings, 
they disappear when the tension of domain wall $\sigma_{\rm wall}$ exceeds that of cosmic string: 
\beq
 \sigma_{\rm wall} (t_2) = \frac{\mu_{\rm string}(t_2)}{t_2}. 
\eeq
Here, the tension of domain wall is given by $\sigma_{\rm wall} (t) \simeq 9.23 m_a (t) F_a^2$. 
The energy density of domain walls is released as axions when they disappear. 
Thus we obtain the axion abundance from the decay of domain walls as 
\beq 
 \Omega_a^{\rm DW} h^2 = 
 (0.004 -0.13)
 \lmk \frac{F_a}{10^{11} \GEV} \rmk^{1.19} 
  \lmk \frac{\Lqcd}{400 \MEV} \rmk. 
\eeq

To sum up the above three contributions, we obtain the total amount of cold axions as 
\beq
 \Omega_{\rm DM} \simeq 
 (0.08 - 0.43) 
 \lmk \frac{F_a}{10^{11} \GEV} \rmk^{1.19} 
  \lmk \frac{\Lqcd}{400 \MEV} \rmk. 
\eeq
This is consistent with the observed DM abundance $\Omega_{\rm DM}^{\rm obs} h^2 \simeq 0.12$ 
when the axion decay constant is given as 
\beq
 F_a \simeq (0.3 - 1.5) \times 10^{11} \GEV, 
 \label{F_a from DM}
\eeq
where we use $\Lqcd = 400 \MEV$.

\subsection{Dark radiation
\label{subsec:dark radiation}}

While cold axions are generated at the time around the QCD phase transition as we calculated in the previous subsection, 
relativistic axions are also produced from thermal plasma at high temperature~\cite{Turner:1986tb}. 
The relevant and model-independent reactions are scatterings between gluons and/or quarks in the thermal plasma, 
whose cross sections are roughly given by $\sigma \sim \alpha_s^3/(8\pi^2 F_a^2)$. 
A detailed calculation in Ref.~\cite{Masso:2002np} reveals that 
those processes keep axions in thermal equilibrium with the thermal plasma 
until the temperature decreases to the following value:%
\footnote{%
Reference~\cite{D'Eramo:2014rna} claims 
that the hadronic processes such as $\pi \pi \leftrightarrow \pi a$ 
are thermal equilibrium below the QCD phase transition for $F_a \le 10^{12} \GEV$. 
However, calculations in Ref.~\cite{Hannestad:2005df} imply that those processes are inefficient for $F_a \gtrsim 10^8 \GEV$. 
In this paper, we follow the latter result which allows us to neglect those reactions for $F_a \sim 10^{11} \GEV$, 
because the estimation by a naive cross section of $\sigma_\pi \sim F_a^{-2}$ 
is inconsistent with the former result by more than four order of magnitude. 
}
\beq
 T_D^{(\rm axion)} \simeq 2 \times 10^{9} \GEV \lmk \frac{F_a}{10^{11} \GEV} \rmk^2. 
\eeq

Next, we consider the thermalization of familons. 
Below the mass scale of $\chi$ the familons interact with the axion through the effective 
interactions 
\beq
 {\cal L}_{eff} \simeq \frac{6}{16 \pi^2 F_a^2 F_f^2} (\del_\mu a)(\del^\mu a) (\del_\nu f)(\del^\nu f) + \dots, 
\eeq
where $F_f$ ($= \mathcal{O}(v_i)$) collectively denotes familon decay constants. 
Thus the familons are produced by the scattering of axions as $a+a \to f+f$. 
The scattering cross sections are roughly given by $\sigma \sim T^6/(F_a^4 F_f^4)$ 
and those scattering processes are decoupled at temperature 
\beq
 T_D^{(\rm familon)} \simeq 10^{9-10} \GEV. 
\eeq

If heavy Higgs doublet bosons have a mass scale $M_H$ less than the reheating temperature, 
the Compton-like scattering processes may keep axions and familons in thermal equilibrium with the thermal plasma. 
This process is decoupled at the temperature around $T \sim M_H$. 
We consider that it is most likely the case, 
since otherwise the required fine tuning of the lightest Higgs boson mass becomes severer. 
However, whether or not these processes are taken into account, 
relativistic axions and familons are decoupled at a temperature above the electroweak phase transition 
and below the PQ phase transition. 
Thus, our conclusions are not altered whether or not we take into account those contributions.

As we show in Sec.~\ref{subsec:leptogenesis} 
the thermal leptogenesis requires $T_{\rm RH} \gtrsim 3 \times 10^{9} \GEV$, 
so that axions and familons are generated from the thermal plasma after reheating. 
Then they are decoupled 
well before the electroweak phase transition 
and contribute to dark radiation in the subsequent cosmological history~\cite{Nakayama:2010vs, Weinberg:2013kea}. 
The amount of dark radiation is conventionally expressed by 
the effective neutrino number $N_{\rm eff}$ defined by 
\beq
 \rho_{\rm rel} = \rho_\gamma \lkk 1 + N_{\rm eff} \frac{7}{8} \lmk \frac{4}{11} \rmk^{4/3} \rkk. 
\eeq
The deviation from the standard value can be calculated from 
\beq
 \Delta N_{\rm eff} = \sum_i \frac{4}{7} \lmk \frac{g_* (T_D^i)}{43/4} \rmk^{-4/3}, 
\eeq
where the summation is taken for all relativistic particles $i$ decoupled at $T_D^i$. 
Using $g_* (T_D) = 106.75$, we obtain $\Delta N_{\rm eff}^{\rm axion} \simeq 0.027$
for the axion contribution to the effective neutrino number. 
Taking into account familons as well as axion, we predict the amount of dark radiation as 
\beq 
 \Delta N_{\rm eff} \simeq 0.24. 
\eeq

An extra background of relativistic particles 
affects the CMB anisotropies by the early integrated Sachs-Wolfe effect. 
The abundances of light elements are also affected by dark radiation through the effect on the expansion rate 
at the BBN epoch. 
The Planck data combined with the observation of helium abundance estimated by Ref.~\cite{Aver:2013wba} 
is consistent with the standard value $N_{\rm eff} = 3.046$ 
but still allows additional radiation component with the constraints~\cite{Planck:2015xua} 
\beq
 N_{\rm eff} = 2.99 \pm 0.39 \qquad (95\% CL). 
\eeq
The measurement of deuterium abundance~\cite{Cooke:2013cba} is consistent with this result. 
On the other hand, Ref.~\cite{Izotov:2014fga} have reported a larger central value 
from the measurement of helium abundance: 
\beq
 N_{\rm eff} = 3.58 \pm 0.40 \qquad (95.4\% CL). 
\eeq
However, it is under discussion whether the uncertainty accurately reflects systematic errors. 
Our result of $\Delta N_{\rm eff} \simeq 0.24$ is consistent with those observations.

Future observations of the CMB fluctuations will measure the effective neutrino number with precisions of 
\beq
 \Delta N_{\rm eff} = 0.0156 \qquad (1 \sigma), 
\eeq
by the ground-based Stage-IV CMB polarization experiment CMB-S4~\cite{Wu:2014hta} (see also Ref.~\cite{Abazajian:2013oma}). 
We conclude that familons and axions will be indirectly observed as dark radiation in the near future. 
The CMB-S4 experiment would indirectly observe axions 
even if dark radiation consists of only axions ($\Delta N_{\rm eff} = 0.027$), 
which is the case of the KSVZ axion model~\cite{Kim:1979if}, for example. 
Let us emphasize that the measurement of the amount of dark radiation 
will give us a probe of the number of NG bosons~\cite{Nakayama:2010vs, Weinberg:2013kea}.

\subsection{Collider signatures
\label{subsec:collider signatures}}

In this subsection, we consider collider signatures predicted in our model. 
In general, NG bosons $f_a$ associated with symmetry breaking interacts with matter fields via 
\beq
 \Lag = \frac{1}{F_{f_a}} \del_\mu f_a 
 j^{a \mu}, 
\eeq
where $j^{a \mu}$ is the broken symmetry current 
and $F_{f}$ is the decay constant. 
In light of the collider experiments, the most relevant interaction is given as%
\footnote{%
Although the familon interactions also lead to flavour changing leptonic decays 
such as $\mu \to e f$ and $\mu \to e \gamma f$, 
these decay modes pose weaker lower bound on $F_f$ than the $K^+$ decay mode does~\cite{Feng:1997tn}. 
We should also note that it is very challenging to improve the limits on these leptonic decay processes 
with the current experimental facilities~\cite{Hirsch:2009ee}. 
}
\beq
  \frac{1}{F_{f_2}} \del_\mu f_2 \bar{s} \gamma^\mu d + h.c., 
 \label{flavour changing interaction}
\eeq
where $F_{f_2}$ is given by $\sqrt{2} \abs{v_1 - v_2}$.%
\footnote{
Here we implicitly assume that the left-handed flavour eigenstates $(d_L, s_L)$ 
interact with familons in the same way as the right-handed ones $(d_R, s_R)$. 
In general, this is not the case 
and we should include the flavour mixing effect. 
In this paper, we neglect the flavour mixing effect for simplicity. 
}
The flavour changing interaction of Eq.~(\ref{flavour changing interaction}) 
leads to an exotic kaon decay process $K^+ \to \pi^+ f$. 
The reaction rate is calculated as~\cite{Feng:1997tn} 
\beq
 \Gamma \lmk K^+ \to \pi^+ f \rmk 
 = 
 \frac{1}{16 \pi} \frac{m_K^3}{F_{f_2}^2} \beta^3 \abs{F_1 (0)}^2, 
\eeq
where $\beta = 1 - m_\pi^2/ m_K^2$. 
The form factor $\la \pi^+ (p') \left\vert \bar{s} \gamma^\mu d \right\vert K^+ (p) \ra = F_1 (q^2) (p+p')^\mu$ 
has a normalization $F_1(0) = 1$ in the limit of exact flavour $SU(3)$ symmetry. 
Collider experiments pose upper bound on the exotic kaon decay rate as~\cite{Artamonov:2009sz}%
\footnote{%
Although Ref.~\cite{Adler:2002hy} have reported the slightly severer upper bound as 
$\Br (K^+ \to \pi^+ f) < 4.5 \times 10^{-11}$, 
we conservatively quote the upper bound derived in the recent paper~\cite{Artamonov:2009sz}. 
If we take the former upper bound, 
the lower bound on $F_{f_2}$ is calculated as $F_{f_2} \gtrsim 9.0 \times 10^{11} \GEV$. 
}
\beq
 \Br (K^+ \to \pi^+ f) < 1.3 \times 10^{-10} 
 \qquad \lmk 90 \% CL \rmk. 
\eeq
This leads to a lower bound on the decay constant: 
\beq
 F_{f_2} \gtrsim 5.3 \times 10^{11} \GEV. 
\eeq
This is marginally consistent with the decay constant $F_a$ determined by the cold axion abundance 
(see Eq.~(\ref{F_a from DM})) 
because 
$F_{f_2}$ ($= \sqrt{2} \abs{v_1 - v_2}$) can be larger than $F_a$ given by Eq.~(\ref{F_a}) 
when one of $v_i$ is smaller than the others.

The recent start of the NA62 experiment at CERN 
is expected to measure the branching ratio of the process $K^+ \to \pi^+ \nu \bar{\nu}$ 
within an accuracy of $10\%$ compared to the SM prediction 
$\Br ( K^+ \to \pi^+ \nu \bar{\nu} )_{\rm SM} = (9.11 \pm 0.72) \times 10^{-11}$~\cite{Buras:2015qea} by 2018~\cite{Anelli:2005ju}. 
This implies that the NA62 experiment will observe a deviation from the SM prediction 
when the decay constant is just above the present lower bound.

\subsection{Neutrino mass and leptogenesis
\label{subsec:leptogenesis}}

In this subsection, 
we introduce right-handed neutrinos $\nubar$ to explain the small masses of left-handed neutrinos 
by the seesaw mechanism~\cite{seesaw, Langacker:1986rj}. 
Since they are charged under $\SUf$ as shown in Table~\ref{table1}, 
their interaction terms are written as 
\beq
 \Lag = - \frac{1}{2} y \chi^{ij} \nubar_i \nubar_j - y' L_i H_u^{ij} \nubar_j + h.c., 
 \label{L for nubar}
\eeq
where $y$ and $y'$ are Yukawa coupling constants. 
After the SSB of $\SUf \times \Upq$, the Lagrangian is reduced to be 
\beq
 \Lag = - \frac{1}{2} M^{ij} _{\nubar} \nubar_i \nubar_j - y^{ij} L_i H_u \nubar_j + h.c., 
\eeq
where $M^{ij} _{\nubar} = y \la \chi^{ij} \ra$. 
Here, we have integrated out heavy Higgs bosons 
and have written the light Higgs $H_u^{ij}$ as $y' H_u^{ij} \equiv y^{ij} H_u$. 
Integrating out $\nubar$, we obtain the mass matrix of left-handed neutrinos via the seesaw mechanism: 
\beq
 m_\nu = - m_D M_{\nubar}^{-1} m_D^T, 
\eeq
where $(m_D)^{ij} \equiv y^{ij} \la H_u \ra$. 
The observed neutrino mass deferences 
$\Delta m_{31}^2 \simeq (7.53 \pm 0.18) \times 10^{-5} \EV^2$ 
and $\Delta m_{32}^2 \simeq (2.44 \pm 0.06) \times 10^{-3} \EV^2$ (normal mass hierarchy)~\cite{Agashe:2014kda} 
imply that $y^{ij} = \mathcal{O}(10^{-3})$ for $M_{\nubar}^{ij} = \mathcal{O}(10^{10}) \GEV$.

The baryon asymmetry of the Universe can also be explained by the thermal leptogenesis~\cite{Fukugita:1986hr}. 
We assume that the reheating temperature is larger than the mass of the lightest 
right-handed neutrino $M_{\nubar_1}$. 
Assuming hierarchical right-handed neutrino masses, $M_{\nubar_{2,3}} \gg M_{\nubar_1}$, 
we can calculate the amount of baryon asymmetry via 
\beq
 Y_b \equiv \frac{n_b}{s} 
 \simeq - \frac{24+4n_H}{66+13n_H} \epsilon \kappa Y^{\rm eq}, 
 \label{Y_b}
\eeq
where $\epsilon$ is the CP-asymmetry parameter in $\nubar_1$ decays, 
and the fractional factor, depending on the number of Higgs doublet $n_H$ ($=2$ in the DFSZ model), 
comes from the sphaleron effect. 
The equilibrium abundance is given by $Y^{\rm eq} = 135 \zeta(3)/(4 \pi^4 g_*)$, 
where $g_*$ is the effective number of spin-degrees of freedom in the thermal plasma. 
The efficiency factor $\kappa$ has to be included to take into account the washout effect and 
the correction of the number density of $\nubar_1$ with respect to the equilibrium value $Y^{\rm eq}$.

The CP-asymmetry parameter is calculated as~\cite{Covi:1996wh}
\beq
 \epsilon &\simeq& - \frac{3}{16\pi} \frac{M_{\nubar_1}}{(y y^\dagger)_{11}} 
 {\rm Im} \lkk \lmk y y^\dagger M_{\nubar}^{-1} y^* y^T \rmk_{11} \rkk \\ 
 &\le& - \frac{3}{16\pi} M_{\nubar_1} \frac{m_{\nu_3}}{ \la H_u \ra^2}, 
 \label{epsilon}
\eeq
where the last inequality is known as the Hamaguchi-Murayama-Yanagida bound~\cite{Hamaguchi:2001gw} 
(see also Ref.~\cite{Davidson:2002qv}). 
It is convenient to define the effective neutrino mass $\tilde{m}_1$ by 
\beq
 \tilde{m}_1 = \sum_i \abs{y_{1 i}}^2 \frac{v^2}{M_1}. 
\eeq
The lightest left-handed neutrino mass $m_{\nubar_1}$ satisfies the inequality of $m_{\nubar_1} \le \tilde{m}_1$. 
The efficiency factor $\kappa$ has been obtained from a detailed calculation in Ref.~\cite{Giudice:2003jh} 
and has been fitted by 
\beq
 \kappa^{-1} \approx \frac{3.3 \times 10^{-3} \EV}{\tilde{m}_1} + \lmk \frac{\tilde{m}_1}{0.55 \times 10^{-3} \EV} \rmk^{1.16}. 
 \label{kappa}
\eeq
The right-handed neutrino $\nubar_1$ cannot be efficiently produced from the thermal plasma 
for the case of $\tilde{m}_1 \ll 10^{-3} \EV$, 
while washout effects are efficient to dilute lepton asymmetry for the case of $\tilde{m}_1 \gg 10^{-3} \EV$. 
These are the reason that the efficiency factor has a maximal value $0.2$ around $\tilde{m}_1 \simeq 10^{-3} \EV$.

The observations of CMB fluctuations imply that the baryon-to-photon ratio is given as~\cite{Agashe:2014kda} 
\beq
 \frac{n_b}{n_\gamma} = (6.05 \pm 0.07) \times 10^{-10}, 
\eeq
where the number density of photon $n_\gamma$ is related to the entropy density as $s = 7.04 n_\gamma$. 
Using Eqs.~(\ref{Y_b}), (\ref{epsilon}), and (\ref{kappa}), 
we can explain the amount of baryon asymmetry by the thermal leptogenesis 
for the case of $T_{\rm RH} \gtrsim M_{\nubar_1} \gtrsim 3 \times 10^{9} \GEV$ 
(see, for a review, Ref.~\cite{Buchmuller:2005eh}). 
Note that the masses of the right-handed neutrinos come from the VEV of $\chi$, 
which is related to the axion decay constant $F_a$. 
The result of $M_{\nubar_1} \gtrsim 3 \times 10^{9} \GEV$ 
is consistent with the result of $F_a = \mathcal{O}(10^{11}) \GEV$ obtained in Sec.~\ref{subsec:DM}. 
The reheating temperature of $T_{\rm RH} \gtrsim 3 \times 10^{9} \GEV$ is also consistent with 
our assumption of 
the thermalized axion and familons, since they are decoupled at temperature around $T \sim 10^9 \GEV$.

\section{Conclusion
\label{sec:conclusion}}

We have proposed a QCD axion model with a $\SUf$ flavour symmetry to solve the domain wall problem by the Lazarides-Shafi mechanism. 
The unwanted discrete PQ symmetry is embedded to the continuous $\SUf$ symmetry to connect the vacua. 
Since the $\SUf$ symmetry is anomalous, it should be global symmetry 
and predicts eight NG bosons called familons in the low energy effective theory. 
We find that familons as well as axions are thermalized and then decoupled 
after the SSB of $\SUf \times \Upq$ symmetry. 
They contribute to dark radiation in the subsequent cosmological history 
and the resulting amount of dark radiation will be detected by future observations of CMB fluctuations. 
Our model also predicts a sizable exotic kaon decay rate, 
which is marginally consistent with the current upper bound. 
We expect that collider experiments would observe excess of kaon decay signals 
such as $K^+ \to \pi^+ f$ in the near future.

\vspace{1cm}

%
\section*{Acknowledgments}
This work is supported by Grant-in-Aid for Scientific research 
from the Ministry of Education, Science, Sports, and Culture
(MEXT), Japan, 
No. 25400248 (M.K.), No. 26104009, and No. 26287039 (T.T.Y), 
World Premier International Research Center Initiative
(WPI Initiative), MEXT, Japan,
and the Program for the Leading Graduate Schools, MEXT, Japan (M.Y.).
M.Y. acknowledges the support by JSPS Research Fellowships for Young Scientists, No.25.8715.
%

\vspace{1cm}




\begin{thebibliography}{90}

\bibitem{Aad:2012tfa} 
  G.~Aad {\it et al.}  [ATLAS Collaboration],
  Phys.\ Lett.\ B {\bf 716}, 1 (2012)
  [arXiv:1207.7214 [hep-ex]].
  
\bibitem{Chatrchyan:2012ufa} 
  S.~Chatrchyan {\it et al.}  [CMS Collaboration],
  Phys.\ Lett.\ B {\bf 716}, 30 (2012)
  [arXiv:1207.7235 [hep-ex]].

\bibitem{Callan:1976je} 
  C.~G.~Callan, Jr., R.~F.~Dashen and D.~J.~Gross,
  Phys.\ Lett.\ B {\bf 63}, 334 (1976).
  
\bibitem{Jackiw:1976pf} 
  R.~Jackiw and C.~Rebbi,
  Phys.\ Rev.\ Lett.\  {\bf 37}, 172 (1976).
  
\bibitem{Baker:2006ts} 
  C.~A.~Baker, D.~D.~Doyle, P.~Geltenbort, K.~Green, M.~G.~D.~van der Grinten, P.~G.~Harris, P.~Iaydjiev and S.~N.~Ivanov {\it et al.},
  Phys.\ Rev.\ Lett.\  {\bf 97}, 131801 (2006)
  [hep-ex/0602020].
  
 
\bibitem{Peccei:1977hh} 
  R.~D.~Peccei and H.~R.~Quinn,
  Phys.\ Rev.\ Lett.\  {\bf 38}, 1440 (1977).

\bibitem{Peccei:1977ur} 
  R.~D.~Peccei and H.~R.~Quinn,
  Phys.\ Rev.\ D {\bf 16}, 1791 (1977).

\bibitem{Kim:1979if} 
  J.~E.~Kim,
  Phys.\ Rev.\ Lett.\  {\bf 43}, 103 (1979);
  M.~A.~Shifman, A.~I.~Vainshtein and V.~I.~Zakharov,
  Nucl.\ Phys.\ B {\bf 166}, 493 (1980). 

 \bibitem{Dine:1981rt} 
  M.~Dine, W.~Fischler and M.~Srednicki,
  Phys.\ Lett.\ B {\bf 104}, 199 (1981); 
  A. P. Zhitnitskii, Sov. J. Phys. {\bf 31} (1980) 260.
  

\bibitem{seesaw} T.~Yangida, in Proceedings of the \textit{``Workshop
on the Unified Theory and the Baryon Number in the Universe''},
Tsukuba, Japan, Feb. 13-14, 1979, edited by O.~Sawada and A.~Sugamoto,
KEK report KEK-79-18  [Conf. Proc., C \textbf{7902131}] 95-99; 
T.~Yanagida, 
Prog.\ Theor.\ Phys.\ \textbf{64}, 1103 (1980); 
M.~Gell-Mann, P.~Ramond and R.~Slansky, in \textit{{''Supergravity''}}
(North-Holland, Amsterdam, 1979) \textit{{eds}}. D.~Z.~Freedom
and P.~van Nieuwenhuizen, Print-80-0576 (CERN) [Conf. Proc., C \textbf{790927}]; see also P.~Minkowski,
Phys.\ Lett.\ B \textbf{67}, 421 (1977).



\bibitem{Fukugita:1986hr} 
  M.~Fukugita and T.~Yanagida,
  Phys.\ Lett.\ B {\bf 174}, 45 (1986).


\bibitem{Weinberg:1977ma} 
  S.~Weinberg,
  Phys.\ Rev.\ Lett.\  {\bf 40}, 223 (1978);
  F.~Wilczek,
  Phys.\ Rev.\ Lett.\  {\bf 40}, 279 (1978).


\bibitem{Preskill:1982cy} 
  J.~Preskill, M.~B.~Wise and F.~Wilczek,
  Phys.\ Lett.\ B {\bf 120}, 127 (1983).
  
\bibitem{Abbott:1982af} 
  L.~F.~Abbott and P.~Sikivie,
  Phys.\ Lett.\ B {\bf 120}, 133 (1983).
  
\bibitem{Dine:1982ah} 
  M.~Dine and W.~Fischler,
  Phys.\ Lett.\ B {\bf 120}, 137 (1983).
    
\bibitem{Kuzmin:1985mm} 
  V.~A.~Kuzmin, V.~A.~Rubakov and M.~E.~Shaposhnikov,
  Phys.\ Lett.\ B {\bf 155}, 36 (1985).
  
  
  
\bibitem{Axenides:1983hj} 
  M.~Axenides, R.~H.~Brandenberger and M.~S.~Turner,
  Phys.\ Lett.\ B {\bf 126}, 178 (1983).
  
\bibitem{Seckel:1985tj} 
  D.~Seckel and M.~S.~Turner,
  Phys.\ Rev.\ D {\bf 32}, 3178 (1985).
  
\bibitem{Turner:1990uz} 
  M.~S.~Turner and F.~Wilczek,
  Phys.\ Rev.\ Lett.\  {\bf 66}, 5 (1991).
  
  
\bibitem{Zeldovich:1974uw} 
  Y.~B.~Zeldovich, I.~Y.~Kobzarev and L.~B.~Okun,
  Zh.\ Eksp.\ Teor.\ Fiz.\  {\bf 67}, 3 (1974)
  [Sov.\ Phys.\ JETP {\bf 40}, 1 (1974)].
  
  
\bibitem{Sikivie:1982qv} 
  P.~Sikivie [ADMX Collaboration],
  Phys.\ Rev.\ Lett.\  {\bf 48}, 1156 (1982).

  
    
\bibitem{Vilenkin:1982ks} 
  A.~Vilenkin and A.~E.~Everett,
  Phys.\ Rev.\ Lett.\  {\bf 48}, 1867 (1982).


\bibitem{Buchmuller:2005eh} 
  W.~Buchmuller, R.~D.~Peccei and T.~Yanagida,
  Ann.\ Rev.\ Nucl.\ Part.\ Sci.\  {\bf 55}, 311 (2005)
  [hep-ph/0502169].
  


\bibitem{Lazarides:1982tw} 
  G.~Lazarides and Q.~Shafi,
  Phys.\ Lett.\ B {\bf 115}, 21 (1982).


\bibitem{Wilczek:1982rv} 
  F.~Wilczek,
  Phys.\ Rev.\ Lett.\  {\bf 49}, 1549 (1982).
  
\bibitem{Reiss:1982sq} 
  D.~B.~Reiss,
  Phys.\ Lett.\ B {\bf 115}, 217 (1982).
  
  
\bibitem{Gelmini:1982zz} 
  G.~B.~Gelmini, S.~Nussinov and T.~Yanagida,
  Nucl.\ Phys.\ B {\bf 219}, 31 (1983).

\bibitem{Kim:1986ax} 
  J.~E.~Kim,
  Phys.\ Rept.\  {\bf 150}, 1 (1987).


\bibitem{Nakayama:2010vs} 
  K.~Nakayama, F.~Takahashi and T.~T.~Yanagida,
  Phys.\ Lett.\ B {\bf 697}, 275 (2011)
  [arXiv:1010.5693 [hep-ph]].
  

\bibitem{Weinberg:2013kea} 
  S.~Weinberg,
  Phys.\ Rev.\ Lett.\  {\bf 110}, no. 24, 241301 (2013)
  [arXiv:1305.1971 [astro-ph.CO]].


  
\bibitem{'tHooft:1976up} 
  G.~'t Hooft,
  Phys.\ Rev.\ Lett.\  {\bf 37}, 8 (1976).
  
  
\bibitem{'tHooft:1976fv} 
  G.~'t Hooft,
  Phys.\ Rev.\ D {\bf 14}, 3432 (1976)
  [Erratum-ibid.\ D {\bf 18}, 2199 (1978)].
  
    
\bibitem{Barr:1982bb} 
  S.~M.~Barr, D.~B.~Reiss and A.~Zee,
  Phys.\ Lett.\ B {\bf 116}, 227 (1982).


 
\bibitem{Davis:1986xc} 
  R.~L.~Davis,
  Phys.\ Lett.\ B {\bf 180}, 225 (1986).
  
  
\bibitem{Lyth:1991bb} 
  D.~H.~Lyth,
  Phys.\ Lett.\ B {\bf 275}, 279 (1992).
  
  

\bibitem{Kawasaki:2014sqa} 
  M.~Kawasaki, K.~Saikawa and T.~Sekiguchi,
  Phys.\ Rev.\ D {\bf 91}, no. 6, 065014 (2015)
  [arXiv:1412.0789 [hep-ph]].



\bibitem{Wantz:2009it} 
  O.~Wantz and E.~P.~S.~Shellard,
  Phys.\ Rev.\ D {\bf 82}, 123508 (2010)
  [arXiv:0910.1066 [astro-ph.CO]].
  
      
\bibitem{Turner:1986tb} 
  M.~S.~Turner,
  Phys.\ Rev.\ Lett.\  {\bf 59}, 2489 (1987)
  [Erratum-ibid.\  {\bf 60}, 1101 (1988)].
  
  
\bibitem{Masso:2002np} 
  E.~Masso, F.~Rota and G.~Zsembinszki,
  Phys.\ Rev.\ D {\bf 66}, 023004 (2002)
  [hep-ph/0203221].

  
\bibitem{D'Eramo:2014rna} 
  F.~D'Eramo, L.~J.~Hall and D.~Pappadopulo,
  JHEP {\bf 1411}, 108 (2014)
  [arXiv:1409.5123 [hep-ph]].
  
\bibitem{Hannestad:2005df} 
  S.~Hannestad, A.~Mirizzi and G.~Raffelt,
  JCAP {\bf 0507}, 002 (2005)
  [hep-ph/0504059].
  
    
\bibitem{Aver:2013wba} 
  E.~Aver, K.~A.~Olive, R.~L.~Porter and E.~D.~Skillman,
  JCAP {\bf 1311}, 017 (2013)
  [arXiv:1309.0047 [astro-ph.CO]].


\bibitem{Planck:2015xua} 
  P.~A.~R.~Ade {\it et al.}  [Planck Collaboration],
  arXiv:1502.01589 [astro-ph.CO].
    
  
  
\bibitem{Cooke:2013cba} 
  R.~Cooke, M.~Pettini, R.~A.~Jorgenson, M.~T.~Murphy and C.~C.~Steidel,
  arXiv:1308.3240 [astro-ph.CO].
  
\bibitem{Izotov:2014fga} 
  Y.~I.~Izotov, T.~X.~Thuan and N.~G.~Guseva,
  Mon.\ Not.\ Roy.\ Astron.\ Soc.\  {\bf 445}, no. 1, 778 (2014)
  [arXiv:1408.6953 [astro-ph.CO]].
  
    
\bibitem{Wu:2014hta} 
  W.~L.~K.~Wu, J.~Errard, C.~Dvorkin, C.~L.~Kuo, A.~T.~Lee, P.~McDonald, A.~Slosar and O.~Zahn,
  Astrophys.\ J.\  {\bf 788}, 138 (2014)
  [arXiv:1402.4108 [astro-ph.CO]].
  

\bibitem{Abazajian:2013oma} 
  K.~N.~Abazajian {\it et al.}  [Topical Conveners: K.N. Abazajian, J.E. Carlstrom, A.T. Lee Collaboration],
  Astropart.\ Phys.\  {\bf 63}, 66 (2015)
  [arXiv:1309.5383 [astro-ph.CO]].

    
  
\bibitem{Feng:1997tn} 
  J.~L.~Feng, T.~Moroi, H.~Murayama and E.~Schnapka,
  Phys.\ Rev.\ D {\bf 57}, 5875 (1998).


\bibitem{Hirsch:2009ee} 
  M.~Hirsch, A.~Vicente, J.~Meyer and W.~Porod,
  Phys.\ Rev.\ D {\bf 79}, 055023 (2009)
  [Erratum-ibid.\ D {\bf 79}, 079901 (2009)]
  [arXiv:0902.0525 [hep-ph]].
  
  
\bibitem{Artamonov:2009sz} 
  A.~V.~Artamonov {\it et al.}  [BNL-E949 Collaboration],
  Phys.\ Rev.\ D {\bf 79}, 092004 (2009)
  [arXiv:0903.0030 [hep-ex]].
  
\bibitem{Adler:2002hy} 
  S.~S.~Adler {\it et al.}  [E787 Collaboration],
  Phys.\ Lett.\ B {\bf 537}, 211 (2002)
  [hep-ex/0201037].
  
  
\bibitem{Buras:2015qea} 
  A.~J.~Buras, D.~Buttazzo, J.~Girrbach-Noe and R.~Knegjens,
  arXiv:1503.02693 [hep-ph].
  
  
\bibitem{Anelli:2005ju} 
  G.~Anelli, A.~Ceccucci, V.~Falaleev, F.~Formenti, A.~Gonidec, B.~Hallgren, P.~Jarron and A.~Kluge {\it et al.},
  CERN-SPSC-2005-013, CERN-SPSC-P-326.
  
  
\bibitem{Langacker:1986rj} 
  P.~Langacker, R.~D.~Peccei and T.~Yanagida,
  Mod.\ Phys.\ Lett.\ A {\bf 1}, 541 (1986).
  

  
\bibitem{Agashe:2014kda} 
  K.~A.~Olive {\it et al.}  [Particle Data Group Collaboration],
  Chin.\ Phys.\ C {\bf 38}, 090001 (2014).
  
  
\bibitem{Covi:1996wh} 
  L.~Covi, E.~Roulet and F.~Vissani,
  Phys.\ Lett.\ B {\bf 384}, 169 (1996)
  [hep-ph/9605319].

\bibitem{Hamaguchi:2001gw} 
  K.~Hamaguchi, H.~Murayama and T.~Yanagida,
  Phys.\ Rev.\ D {\bf 65}, 043512 (2002)
  [hep-ph/0109030].

\bibitem{Davidson:2002qv} 
  S.~Davidson and A.~Ibarra,
  Phys.\ Lett.\ B {\bf 535}, 25 (2002)
  [hep-ph/0202239].

\bibitem{Giudice:2003jh} 
  G.~F.~Giudice, A.~Notari, M.~Raidal, A.~Riotto and A.~Strumia,
  Nucl.\ Phys.\ B {\bf 685}, 89 (2004)
  [hep-ph/0310123].

  
\end{thebibliography}
\end{document}